\documentclass[twocolumn,showpacs,amsmath,amssymb]{revtex4-1}
\usepackage{graphicx}
\begin{document}
\title{\Large \bf More on  New Massive Gravity: Exact Solutions}
\author{\large Haji Ahmedov and Alikram N. Aliev}
\address{Feza G\"ursey Institute, \c Cengelk\" oy, 34684   Istanbul, Turkey}
\date{\today}

\begin{abstract}

We give a novel description of the recently proposed theory of new  massive gravity (NMG) in three dimensions. We  show that in terms of a Dirac type differential operator acting on the traceless Ricci tensor, the field equations of the theory reduce to the massive Klein-Gordon type equation with a curvature-squared source term and to a constraint equation. Under a certain relation between the source tensor and the traceless Ricci tensor, fulfilled for constant scalar curvature,  the field equations of topologically massive gravity (TMG) can be thought of as the ``square-root" of  the massive Klein-Gordon type equation. Using this fact, we establish  a simple framework for mapping all known algebraic types D and N  solutions of TMG into NMG. We also present new exact solutions of algebraic types D and N which are  only inherent in NMG.

\end{abstract}


\maketitle

Continuing search for a consistent theory of quantum gravity has stimulated many investigators to explore gravity models in three dimensions, where one may hope to have less austere ultraviolet (UV) divergences in perturbation theory. Ordinary general relativity (GR) in three dimensions becomes dynamically trivial as it does not propagate any physical degrees of freedom \cite{djh}.  Remarkably, this problem can  be cured  by a particular extension of GR in three dimensions. There exist two popular approaches to such an extension: (i) the Einstein-Hilbert (EH)  action is supplemented with a  {\it parity-violating} gravitational Chern-Simons term. The resulting theory is known as {\it topologically massive gravity} \cite{djt, deser}. This is a dynamical theory of gravity with a single propagating  massive graviton of helicity  $+$2 or $-$2. The theory turns out to be unitary  for the ``wrong" sign (opposite to that in the four dimensional case) of the EH term in the total action \cite{djt}, (ii) the EH action is extended  by adding  a particular higher-derivative correction term to it. This gives rise to a  novel theory of three-dimensional massive gravity, known as {\it new massive gravity} \cite{bht}.

In contrast to TMG, new massive gravity is a {\it parity-preserving} theory. At the linearized level it becomes  equivalent to the Pauli-Fierz theory  for a free massive graviton in three dimensions, thereby sharing its unitary property \cite{bht} (see also \cite{naka, deser1, tekin}). This allows one to think of NMG  as an interacting and generally covariant extension of the Pauli-Fierz theory. The linearized  equations of NMG for metric perturbations are of fourth-order in derivatives, but they have the required helicity content, describing the physical massive gravitons of helicities $\pm2 $ . The theory  is also  power-counting UV finite \cite{deser1}, which is a  consequence of the renormalizability of its ``cousins" in four dimensions \cite{stelle}. It was shown that   Banados-Teitelboim-Zanelli (BTZ) \cite{btz} and  warped anti-de Sitter (AdS$_3$) black hole solutions of TMG  persist in cosmological NMG  as well \cite{bht, clement}. Some  AdS$_3$-wave  solutions of NMG, which are  counterparts  of those in TMG, were studied in \cite{giri}.

An analysis of  NMG in the context of the AdS$_3$/CFT$_2$  correspondence  reveals the bulk/boundary unitarity conflict: the unitarity  in the bulk implies a negative central charge for  the  boundary CFT$_2$ \cite{berg, liu1,liu2}. In TMG this conflict is resolved, with the ``right" sign EH term in the action, at a ``chiral" point, at which the Compton wavelength of the massive graviton becomes equal to the radius of the AdS$_3$ space \cite{strom}. However, for NMG a similar strategy of the chiral point shows that both the energy of massive bulk modes and the central charges of the dual CFT$_2$ vanish \cite{liu1,liu2}. Thus, the theory becomes trivial at this point under the standard Brown-Henneaux boundary conditions \cite{brown} (see also \cite{daniel1, maloney, daniel2} for some further investigations). In light of all these developments, it becomes of great importance to  undertake an exhaustive study of exact solutions to NMG in hope of finding  a stable vacuum  for a consistent theory of quantum gravity in three dimensions.

The purpose of this Letter is to give a novel description of NMG in terms of a Dirac type differential operator acting on the traceless Ricci tensor and  to establish a simple framework for mapping  all  known Petrov-Serge types D and N  exact solutions of TMG into  NMG, thereby providing a large class of new  exact solutions to NMG. Furthermore, we present new nontrivial solutions of Bianchi types $ VI_{0} $ and $ VII_{0} $  as well as a new type N solution, which
are only inherent in NMG.

We begin by recalling the field equations of NMG \cite{bht}
\begin{eqnarray}
R_{\mu\nu} - \frac{1}{2}\, R g_{\mu\nu} + \lambda g_{\mu\nu} -\frac{1}{2 m^2}\, K_{\mu\nu}& = & 0\,,
\label{nmgfieldeqs1}
\end{eqnarray}
where $ R= g^{\mu\nu}R_{\mu\nu} $ is the three-dimensional Ricci scalar, $\,\lambda $ is a cosmological parameter, $ m $ is a mass parameter  and $ K_{\mu\nu} $ is a symmetric and  covariantly conserved tensor given by
\begin{eqnarray}
K_{\mu\nu} &=& 2 \nabla^2 R_{\mu\nu} -  \frac{1}{2}\left(\nabla_{\mu} \nabla_{\nu} R + g_{\mu\nu} \nabla^2 R \right) -8 R_{\mu}^{\,\,\alpha} R_{\nu \alpha}
\nonumber \\ [2mm] &&
+\frac{9}{2} R  R_{\mu\nu} + g_{\mu\nu}\left(3 R_{\alpha \beta}R^{\alpha\beta}-\frac{13}{8} R^2\right).
\label{ktensor}
\end{eqnarray}
Here $ \nabla_{\mu} $ is the covariant derivative operator with respect to the spacetime metric and $ \nabla^2 =  \nabla_{\mu}\nabla^{\mu} $. The trace of equation (\ref{nmgfieldeqs1}) gives
\begin{eqnarray}
R_{\mu\nu}R^{\mu\nu}- \frac{3}{8} \,R^2 + m^2 R &=& 6 m^2 \lambda\,.
\label{trfeqs}
\end{eqnarray}

We also recall the field equations of TMG \cite{djt,deser},
\begin{eqnarray}
R_{\mu\nu} - \frac{1}{2}\, R g_{\mu\nu}  + \Lambda g_{\mu\nu} + \frac{1}{\mu}\, C_{\mu\nu}=0\,,
\label{tmgfieldeqs1}
\end{eqnarray}
where $ \Lambda $ is the cosmological constant, $ \mu $ is a mass parameter and $ C_{\mu\nu} $ is the Cotton tensor,
\begin{eqnarray}
C_{\mu\nu} &= & {\epsilon_{\mu}}^{\alpha\beta}\nabla_{\alpha}\left(R_{\nu\beta} - \frac{1}{4}\, g_{\nu\beta} R\right)\,,
\label{cotton}
\end{eqnarray}
which is  a symmetric, traceless and covariantly conserved quantity. Here $\epsilon_{\mu\alpha\beta} $ is the Levi-Civita tensor  given by the relation $\epsilon_{\mu\alpha\beta}= \sqrt{-g} \,\varepsilon_{\mu\alpha\beta} $. We use the convention $\varepsilon_{012}=1 $.

The trace of equation (\ref{tmgfieldeqs1}) yields
\begin{eqnarray}
R=6\Lambda \,.
\label{tmgtrace}
\end{eqnarray}
Using the traceless Ricci tensor,
\begin{eqnarray}
S_{\mu\nu}= R_{\mu\nu}-\frac{1}{3}\,g_{\mu\nu} R\,,
\label{trlessricci}
\end{eqnarray}
one can pass to the alternative form of the field equations in (\ref{tmgfieldeqs1}). We have
\begin{eqnarray}
S_{\mu\nu} + \frac{1}{\mu}\, C_{\mu\nu}=0\,.
\label{tmgfieldeqs2}
\end{eqnarray}

Next, we introduce a first-order differential operator  $ {D\hskip -.25truecm \slash} \,$, whose action on a symmetric tensor $ \Phi_{\mu \nu} $ is given by
\begin{eqnarray}
{D\hskip -.25truecm \slash}\,\Phi_{\mu\nu} &= & \frac{1}{2}\left( {\epsilon_{\mu}}^{\alpha\beta}\nabla_{\beta} \Phi_{\nu\alpha} + {\epsilon_{\nu}}^{\alpha\beta}\nabla_{\beta} \Phi_{\mu\alpha} \right)\,.
\label{doper1}
\end{eqnarray}
It is not difficult to show that this expression  reduces to the form
\begin{eqnarray}
{D\hskip -.25truecm \slash\,}\Phi_{\mu\nu} &= & {\epsilon_{\mu}}^{\alpha\beta}\nabla_{\beta} \Phi_{\nu\alpha} \,, \label{doper2}
\end{eqnarray}
provided that
\begin{eqnarray}
\nabla^{\nu}\Phi_{\mu\nu} &= & \nabla_{\mu}\Phi \,,
\label{cond1}
\end{eqnarray}
where $ \Phi $ is the trace of the tensor  $ \Phi_{\mu\nu} $. Choosing this tensor  as
\begin{eqnarray}
\Phi_{\mu\nu}= R_{\mu\nu}-\frac{1}{4}\,g_{\mu\nu} R\,,
\label{phitensor}
\end{eqnarray}
for which condition (\ref{cond1}) is fulfilled, and comparing equations (\ref{cotton}), (\ref{doper1}) and (\ref{doper2})  we find that
\begin{eqnarray}
C_{\mu\nu} &= & -{D\hskip -.25truecm \slash\, }  \Phi_{\mu\nu}= - {D\hskip -.25truecm \slash\, } S_{\mu\nu}\,.
\label{cot1}
\end{eqnarray}
Taking this into account in equation (\ref{tmgfieldeqs2}), we obtain
\begin{eqnarray}
{D\hskip -.25truecm \slash\,} S_{\mu\nu} & = &  \mu S_{\mu\nu}\,.
\label{tmgdirac}
\end{eqnarray}
It is interesting to note that this equation closely resembles
the Dirac equation $ {D\hskip -.25truecm \slash\, }\Psi_{A}=\gamma^{\,\mu B}_{A} \nabla_{\mu} \Psi_B = \mu \Psi_{A} $.  Here the Dirac type operator $ {D\hskip -.25truecm \slash\,} $ acts on the  traceless Ricci tensor, $ {D\hskip -.25truecm \slash\,} S _{\mu\nu} =D_{(\mu\nu)}^{\,\,\,(\alpha\beta) \rho} \nabla_{\rho} S_{\alpha\beta }$. Clearly, the action of $ {D\hskip -.25truecm \slash\,} $ on equation (\ref{tmgdirac}) leads to the second-order Klein-Gordon type equation
\begin{eqnarray}
\left({D\hskip -.25truecm \slash\,}^2  - \mu^2 \right)S_{\mu\nu} & = &  0\,.
\label{tmgkg}
\end{eqnarray}

We now rewrite the field equations of  NMG given in (\ref {nmgfieldeqs1}) in terms of  the traceless Ricci tensor $ S_{\mu\nu} $ and the Dirac type operator $ {D\hskip -.25truecm \slash\,} $  defined above. Using the fact that the Cotton tensor (\ref{cotton}) satisfies  relation (\ref{cond1}) and taking into account equations (\ref{doper2}) and (\ref{cot1}), we obtain that
\begin{eqnarray}
{D\hskip -.25truecm \slash\,}^2  S_{\mu\nu} &=& -{\epsilon_{\mu}}^{\alpha\beta}\nabla_{\beta} C_{\nu\alpha}\,.
\label{nmgdirac1}
\end{eqnarray}
It is  straightforward to show that this expression, with equations (\ref{cotton}) and  (\ref{phitensor}) in mind, can be written in the form
\begin{eqnarray}
{D\hskip -.25truecm \slash\,}^2  S_{\mu\nu} &=& \nabla^2 \Phi_{\mu\nu}- \nabla^{\rho}\nabla_{\nu}\Phi_{\mu\rho}\,.
\label{nmgdirac2}
\end{eqnarray}
This equation can be transformed further by using  condition (\ref{cond1}) and the standard relation between the Riemann and  the Ricci tensors in three dimensions
\begin{eqnarray}
R_{\mu \nu \alpha \beta} & = & 2\left(R_{\mu [\alpha}g_{\beta]\nu} - \,R_{\nu [\alpha} g_{\beta]\mu}\nonumber
\right. \\[2mm]  & & \left.
-\frac{R}{2}\, g_{\mu[\alpha}g_{\beta]\nu}\right)\,.
\label{rimtoric}
\end{eqnarray}
Using then the resulting expression in equation (\ref{nmgfieldeqs1}), we reduce it into the form of the massive Klein-Gordon type equation  with a curvature-squared source term. Thus, we obtain
\begin{eqnarray}
\left({D\hskip -.25truecm \slash\,}^2  - m^2 \right) S_{\mu\nu} & = & T_{\mu\nu}\,,
\label{nmgfieldeqs2}
\end{eqnarray}
where the traceless source term  is given  by
\begin{eqnarray}
T_{\mu\nu}& = &  S_{\mu\rho}S^{\rho}_{\,\nu}- \frac{R}{12}\,S_{\mu\nu}- \frac{1}{3}\,g_{\mu\nu} S_{\alpha\beta}S^{\alpha\beta}\,.
\label{source}
\end{eqnarray}
Meanwhile, the trace equation in (\ref{trfeqs})  takes the form
\begin{eqnarray}
S_{\mu\nu}S^{\mu\nu} + m^2 R - \frac{R^2}{24} & = &  6 m^2 \lambda\,.
\label{newtrfeqs}
\end{eqnarray}
We note that in this description, the field equations NMG reduce to two independent equations (\ref{nmgfieldeqs2}) and (\ref{newtrfeqs}), where the latter equation can be thought of as a constraint. In the canonical description, equation (\ref{trfeqs}) is a consequence of (\ref{nmgfieldeqs1}). As we shall see below, this description of NMG greatly simplifies the search for exact solutions to the theory.

Let us now suppose that the source tensor in (\ref{source})  fulfils the  relation
\begin{eqnarray}
T_{\mu\nu} & = & \kappa  S_{\mu\nu}\,,
\label{geocond}
\end{eqnarray}
where $ \kappa  $ is  a function of the scalar curvature. Then equation (\ref{nmgfieldeqs2}) takes the form
\begin{eqnarray}
\left({D\hskip -.25truecm \slash\,}{^2}-  \mu^2\right)S_{\mu\nu} & = &  0\,,
\label{kgfinal}
\end{eqnarray}
where
\begin{eqnarray}
 \mu^2 & = &  m^2 + \kappa\,.
\label{kgfinal}
\end{eqnarray}
Comparing  this equation with that in (\ref{tmgkg}), we see that  for spacetimes with constant scalar curvature, $ \kappa=const  $, they become equivalent to each other. That is, in the case under consideration, the field equations of TMG  in (\ref{tmgdirac}) can be thought of as the square-root of those  of NMG given in (\ref{kgfinal}). This fact has a striking consequence for mapping all known  algebraic types D and N  exact solutions of TMG into NMG.

We first focus on algebraic type  D solutions. It turns out  that {\it  for every nontrivial algebraic type  D solution of TMG  there exist two  inequivalent type  D solutions to NMG, provided that the solution parameters  are related by}
\begin{eqnarray}
\label{unimas11}
\mu^2 &=&\frac{9 m^2}{7}\left( 2 \pm \,\frac{ \sqrt{5+7\lambda/m^2}}{\sqrt{3}}\right),\\[2mm]
\Lambda & = & - \frac{2 m^2}{21}\left(13 \pm  \,\frac{10\, \sqrt{5+7\lambda/m^2}}{\sqrt{3}}\right).
\label{unicos22}
\end{eqnarray}
The proof of this statement is straightforward.  We recall that type D spacetimes in three dimensions are split into types $ D_t $ and $ D_s $, depending on whether the eigenvector of  $ S^{\mu}_{\,\,\nu} $ is timelike or spacelike (see, for instance, \cite{chow}). For type $ D_t $ spacetimes the canonical form of the traceless Ricci tensor $ S_{\mu\nu} $ is given by
\begin{eqnarray}
S_{\mu\nu} & = & p \left(g_{\mu\nu} +3 t_{\mu} t_{\nu}\right)\,,
\label{riccican}
\end{eqnarray}
where $ p $ is a scalar function  and $  t_{\mu} $ is a timelike vector normalized as $  t_{\mu}  t^{\mu}=-1 $. In works  \cite{chow, gurses}, it was shown that when $ p $ is constant and $  t_{\mu} $ is a Killing vector, obeying the equation
\begin{eqnarray}
\nabla_{\nu}k_{\mu}& = &\frac{\mu}{3}\,\epsilon_{\mu\nu \sigma}k^{\sigma}\,,
\label{killing1}
\end{eqnarray}
the field equations  of TMG  are solved, fixing the value of $ p $.
With this in mind, substituting (\ref{riccican}) in (\ref{geocond})
we find that
\begin{eqnarray}
\kappa &=& - p -\frac{R}{12}\,.
\label{kappa1}
\end{eqnarray}
For any vector with  $ \nabla_{\mu} k^{\mu} =0 $, we have
\begin{eqnarray}
\nabla^{\mu} \nabla_{\nu}k_{\mu}& = & R_{\nu}^{\,\,\sigma} k_{\sigma}= -\left(2 p -
\frac{R}{3}\right) k_{\nu}\,,
\label{ricdef}
\end{eqnarray}
where in the last step we have used equations (\ref{trlessricci}) and (\ref{riccican}).  From  equations (\ref{killing1}) and  (\ref{ricdef}) it follows that
\begin{eqnarray}
6p &= &\frac{2}{3}\,\mu^2 + R\,.
\label{al1}
\end{eqnarray}
Combining  now this equation with those in (\ref{kgfinal}) and  (\ref{kappa1}), and taking into account equation (\ref{tmgtrace}),  we  obtain
\begin{eqnarray}
m^2 -\frac{10}{9}\,\mu^2=\frac{3}{2}\,\Lambda\,.
\label{uni11}
\end{eqnarray}
Meanwhile, substitution of (\ref{riccican}) into equation (\ref{newtrfeqs}), with equations (\ref{tmgtrace}) and (\ref{al1}) in mind, yields
\begin{eqnarray}
\frac{1}{6}\left(\frac{2}{3}\,\mu^2 + 6 \Lambda \right)^2 + 6 m^2 \Lambda - \frac{3}{2}\,\Lambda^2 & = &  6 m^2 \lambda\,.
\label{uni22}
\end{eqnarray}
Solving algebraic equations (\ref{uni11}) and (\ref{uni22}), we arrive at the relations given in (\ref{unimas11}) and (\ref{unicos22}). A similar analysis shows that these relations remain unchanged for  type $ D_s $ spacetimes.

We now show that {\it  every algebraic type  N  spacetime of TMG  provides two inequivalent type N solutions to NMG and the parameters of the solutions are adjusted  according to the relations}
\begin{eqnarray}
\label{fnuni11}
\mu^2 &=& \mp \,m^2 \sqrt{1-\lambda/m^2}\,,\\[2mm]
\Lambda & = & 2 m^2\left(1\pm \sqrt{1-\lambda/m^2}\right).
\label{nnuni22}
\end{eqnarray}
For TMG spacetimes of type N, the canonical form of the  traceless Ricci tensor is given by
\begin{eqnarray}
S_{\mu\nu}= l_{\mu}l_{\nu}\,,
\label{Nricci}
\end{eqnarray}
where  $ l_{\mu} $ is a null vector \cite{gibbons}. Substitution of this tensor into equations (\ref{newtrfeqs}) and (\ref{geocond}) yields
\begin{eqnarray}
\Lambda (4 m^2 -\Lambda) &=& 4 m^2\lambda\,
\label{trnuni11}
\end{eqnarray}
and $ \kappa =  -\Lambda/2 $, respectively. Hence,  equation (\ref{kgfinal}) takes the form
\begin{eqnarray}
m^2 -\mu^2 &=& \frac{1}{2}\,\Lambda\,,
\label{nuni11}
\end{eqnarray}
Again, solving algebraic equations (\ref{trnuni11}) and (\ref{nuni11})  we obtain the relations given in (\ref{fnuni11}) and (\ref{nnuni22}).

We now turn to exact type D solutions of NMG gravity which do not have their counterparts in TMG. In TMG, as seen from equation (\ref{tmgdirac}), the action of the operator  $ {D\hskip -.25truecm \slash\,} $  on the traceless Ricci tensor $ S_{\mu\nu} $ results in the same type geometry. This not the case for NMG in general, where the same operation leads to ``intermediate" geometries as well. However, the secondary action of $ {D\hskip -.25truecm \slash\,} $ restores the original algebraic type D geometry \cite{ah}. Remarkably, this fact provides us with a new class of nontrivial type D solutions that are only inherent in  NMG. We present below two simple examples of such solutions.

The first solution, for the cosmological parameter $  \lambda= m^2/5 $, is given by
\begin{eqnarray}
ds^2 &=& -dt^2 + e^{2 \sqrt{2/5}\,m t}\,dx^2 + e^{-2\sqrt{2/5}\,m t}\, dy^2\,.
\label{metric1}
\end{eqnarray}
This metric admits a three-parameter group of motions and the associated Killing vectors
\begin{eqnarray}
\xi_1 &= & \partial_x\,,~~ \xi_2 =  \partial_y\,, ~~ \xi_3 =  - \partial_t +  \sqrt{2/5}\,m \left(x \partial_x - y \partial_y\right)\nonumber \\
\label{killings6}
\end{eqnarray}
satisfy the Lie algebra
\begin{eqnarray}
[\xi_1 , \xi_2]  &=&  0\,,~ [\xi_1 ,\xi_3]  = \sqrt{2/5}\,m \xi_1 \,,~ [\xi_2 , \xi_3]  = - \sqrt{2/5}\,m  \xi_2\,.\nonumber\\
\end{eqnarray}
That is, the metric in (\ref{metric1}) corresponds to  a homogeneous anisotropic spacetime of Bianchi type $ VI_{0} $, or with $ E(1,1) $  symmetry.

The second solution  is of  a homogeneous anisotropic spacetime of Bianchi type $ VII_{0} $  with a three-parameter group of motions. We have
\begin{eqnarray}
ds^2 &=& \cos (2 \sqrt{2/5}\,m x) \left(-dt^2+ dy^2\right)+ dx^2\nonumber \\
&&+ 2\sin(2 \sqrt{2/5}\,m x) dt dy \,,
\label{metric2}
\end{eqnarray}
for  the cosmological parameter $  \lambda= m^2/5 $,  and  with  the Killing vectors
\begin{equation}
\xi_1 =  \partial_t\,,~~ \xi_2 =  \partial_y\,, ~~ \xi_3 =   \partial_x - \sqrt{2/5}\,m \left(t \partial_y - y \partial_t\right),
\label{killings7}
\end{equation}
satisfying  the  Lie algebra
\begin{eqnarray}
[\xi_1 , \xi_2] & = & 0\,, ~ [\xi_1 ,\xi_3]  = -\sqrt{2/5}\,m \xi_2  \,,~ [\xi_2 , \xi_3]  =  \sqrt{2/5}\,m  \xi_1 \nonumber \\
\end{eqnarray}
of $ E(2) $  symmetry.

Similarly, the novel description of NMG enables us to obtain  new nontrival  type  N  solutions that  are  inherent in NMG \cite{ah}. A remarkably simple example with $\lambda= -\nu^2 (1+\nu^2/4m^2) $ is given by
 \begin{eqnarray}
ds^2 &=& d\rho^2 + 2\cosh^2(\nu\rho) du dv  + \cosh(\nu\rho) \left[\cosh(\mu\rho) f(u) \right. \nonumber \\[2mm]  & & \left.
- v^2 \nu^2 \cosh(\nu\rho)\right] du^2\,,
\label{metric3}
\end{eqnarray}
where $ f(u) $ is an arbitrary function and $ \nu=\sqrt{-\Lambda}\, $. This metric does not admit the null Killing vector and belongs to a class of Kundt spacetimes. However, when $ \Lambda \rightarrow 0 $, the Killing isometry appears and $ \partial_v $ becomes a covariantly constant null Killing vector. That is, the resulting metric
 \begin{eqnarray}
ds^2 &=& d\rho^2 + 2 du dv  + \cosh(m\rho) f(u) du^2\,,
\label{metric4}
\end{eqnarray}
represents  pp-waves solution being the  limiting case of AdS pp-waves, found earlier in \cite{giri}.

In summary, the results presented in this Letter are of interest for several reasons: first of all, we have given a novel description of NMG in terms of a Dirac type differential operator acting on the traceless Ricci tensor. Namely, we have shown that  the field equations of NMG  acquire the form of the massive Klein-Gordon type equation with curvature-squared source term, which is also accompanied by  a constraint equation. It turns out that {\it such a description has a striking consequence, greatly simplifying the search for exact solutions to NMG}.

We have found that under  a certain  relation between the source tensor and the traceless Ricci tensor, fulfilled for constant scalar curvature, the field equation of TMG can be thought of as the square-root of  the massive Klein-Gordon type equation of NMG. This fact enabled us to establish a  simple framework, involving in essence an algebraic procedure, for mapping all  known algebraic types D and N  exact solutions of TMG into NMG, thereby providing a large class new nontrivial  exact solutions to NMG. Furthermore, the novel description of NMG  also provides  a large class of types D and N  new exact solutions which  do not have their counterparts in TMG.  As the most simple examples of these spacetimes, we have presented new solutions of  Bianchi types $ VI_{0} $ and $ VII_{0} $ as well as a new type N solution. An exhaustive investigation of all types D and N exact solutions  to  NMG will be given in  our forthcoming paper \cite{ah}.

\end{document}